\documentclass[a4paper,11pt]{article}

\usepackage{pos}
\usepackage{lipsum} 

\title{Indirect dark matter search in the Galactic Centre with IceCube}

\ShortTitle{Indirect dark matter search in the Galactic Centre with IceCube}

\author{The IceCube Collaboration \\{\normalsize \normalfont(a complete list of authors can be found at the end of the proceedings)}\\}

\emailAdd{tchau@icecube.wisc.edu}
\emailAdd{juanan.aguilar@icecube.wisc.edu}

\abstract{
It is assumed that dark matter can annihilate or decay into Standard Model particles which then can produce a neutrino flux detectable at IceCube. Such a signal can be emitted from the Galactic Center thanks to the high density of dark matter abundance being gravitationally captured. This analysis aims at searching for neutrino signals from dark matter annihilation and decay in the Galactic Center using $\sim$9 years of IceCube-DeepCore data with an optimized selection for low energy. In this contribution, we present the sensitivities on the thermally averaged dark matter self-annihilation cross-section for dark matter masses ranging from 5 GeV up to 8 TeV.

\vspace{4mm}
{\bfseries Corresponding authors:}
Nhan Chau$^{1*}$, $^{1}$Juan A. Aguilar\\
{$^{1}$ \itshape Université Libre de Bruxelles}\\
$^*$ Presenter

\ConferenceLogo{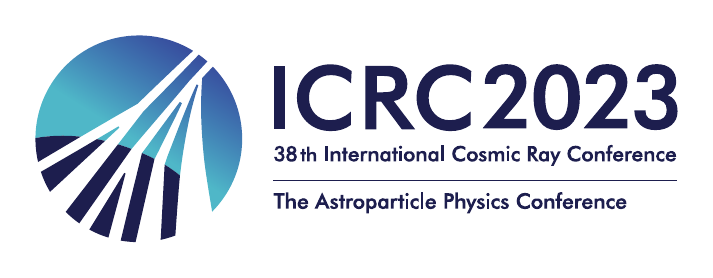}

\FullConference{The 38th International Cosmic Ray Conference (ICRC2023)\\ 26 July -- 3 August, 2023\\ Nagoya, Japan}
}

\begin{document}
\maketitle

\section{Introduction}
The search for dark matter (DM) has been one of the most captivating fields in fundamental physics. While the existence of dark matter has been well established by multiple astrophysical observations \cite{RevModPhys.90.045002, Planck:2013pxb}, the nature of it is still an unresolved question. One popular hypothesis is the `particle solution', where dark matter is assumed to be long-lived or stable particles. This is supported by multiple Beyond Standard Model (BSM) theories that propose a wide class of DM candidates. In many of these models, dark matter candidates can have weak coupling to Standard Model (SM) particles which then allows them to decay or annihilate into detectable SM particles \cite{sym12101648, RevModPhys.93.035007}.

In the indirect search for dark matter, one investigates the presence of unconventional fluxes of Standard Model (SM) particles generated through dark matter decay or annihilation. This signal becomes detectable when there is a substantial accumulation of dark matter, gravitationally captured by massive astrophysical entities. The Galactic Centre (GC) stands out as a particularly promising source for such investigations.

In this analysis, we present a search with IceCube \cite{Aartsen:2016nxy} for the neutrino signal from the Galactic Centre as secondary or primary products from dark matter decay and annihilation into Standard Model particles. Previous studies of IceCube have performed such searches with cascades \cite{IceCube:2016oqp} and tracks \cite{IceCube:2017rdn} only using directional information. In a recent IceCube analysis \cite{IceCube:2023ies}, both the directional and energy distributions of the events are included from the data sample consisting mostly of cascade events.  In this work, we incorporate both track and cascade events, as well as considering both directional and energy distributions. The analysis has also applied the latest optimized selection criteria specifically designed for sub-TeV events. The objective is to improve the IceCube limit for dark matter masses ranging from 5 GeV up to 8 TeV using the DeepCore data from 2012-2022. DeepCore \cite{IceCube:2011ucd} is a subarray of IceCube, characterized by a higher concentration of digital optical modules compared to the remaining of IceCube. The purpose of this subarray is to lower the neutrino energy threshold to energies as low as about 10 GeV.
\section{Event Selection}
Since the analysis focuses mostly on sub-TeV DM masses, we use the DeepCore data with energies ranging from 1 GeV - 1 TeV. At this energy range, the primary background originates from atmospheric muons and neutrinos generated through interactions of cosmic rays in the upper atmosphere. In the case of particles from the Northern hemisphere, the Earth acts as a natural shield, effectively reducing the impact of atmospheric muons as background noise. However, for sources in the southern sky, such as the Galactic Centre, a specific protocol of using a veto is necessary to distinguish atmospheric muons from other signals.

In this analysis, we employ a well-established event selection known as oscNext \cite{IceCube:2023ewe}, which is tailored for precise measurements of atmospheric neutrino oscillations. This selection focuses on the DeepCore sub-detector within IceCube while using the remaining parts of the detector as a veto. The oscNext event selection currently encompasses DeepCore events recorded between 2012 and 2022, providing a total livetime of 9.3 years. 

In the context of the oscNext event selection, the standard oscillation analyses only select up-going events within an energy range of up to 500 GeV to fulfill the oscillation studies. However, in this analysis, we include all direction events because the Galactic Centre is above the horizon for IceCube. Additionally, we have extended the energy range from 1 GeV to 1 TeV to accommodate the investigation of $\sim$TeV dark matter masses.

\section{Dark Matter signal from the Galactic Centre}

The incoming differential flux of neutrinos from dark matter self-annihilation or decay in the Galactic Centre can be computed as

\begin{equation}\label{eq:trueflux}  
 \frac{\mathrm{d}\phi_{\nu}}{\mathrm{d}E_{\nu}} = \left\{
\begin{aligned}
&\frac{1}{4\pi} \, \frac{\langle \sigma \upsilon \rangle}{2 \, m_{\mathrm{DM}}^2} \; \frac{\mathrm{d}N_{\nu}}{\mathrm{d}E_{\nu}} \; J \qquad \text{for annihilation} \, ,\\
&\frac{1}{4\pi} \, \frac{1}{ m_{\mathrm{DM}} \tau} \; \frac{\mathrm{d}N_{\nu}}{\mathrm{d}E_{\nu}} \; J \qquad \text{for decay} \, ,
\end{aligned}
\right.
\end{equation}
with $\langle \sigma \upsilon \rangle $ being the thermally-averaged self-annihilation cross-section, $\tau$ being the DM decay lifetime, and $m_{\mathrm{DM}}$ being the mass of the dark matter particles. The flux also depends on the differential number of neutrinos per DM annihilation/decay, $\mathrm{d}N_{\mathrm{\nu}}/\mathrm{d}E_{\mathrm{\nu}}$, where the assumption of a 100\% branching ratio into either primary channels of $W^+W^-, \mu^+ \mu^-, \tau^+ \tau^-, b\overline{b}, \nu_e \overline{\nu}_e, \nu_\mu \overline{\nu}_\mu, \nu_\tau \overline{\nu}_\tau$ is adopted. The entity $J$ referred to as J-factor encodes the shape of the dark matter halo. While the neutrino spectrum $\mathrm{d}N_{\mathrm{\nu}}/\mathrm{d}E_{\mathrm{\nu}}$ governs the energy dependence of the expected signal, the J-factor impacts the spatial distribution of such flux.

\begin{figure}
    \centering
    \includegraphics[width=0.48\textwidth]{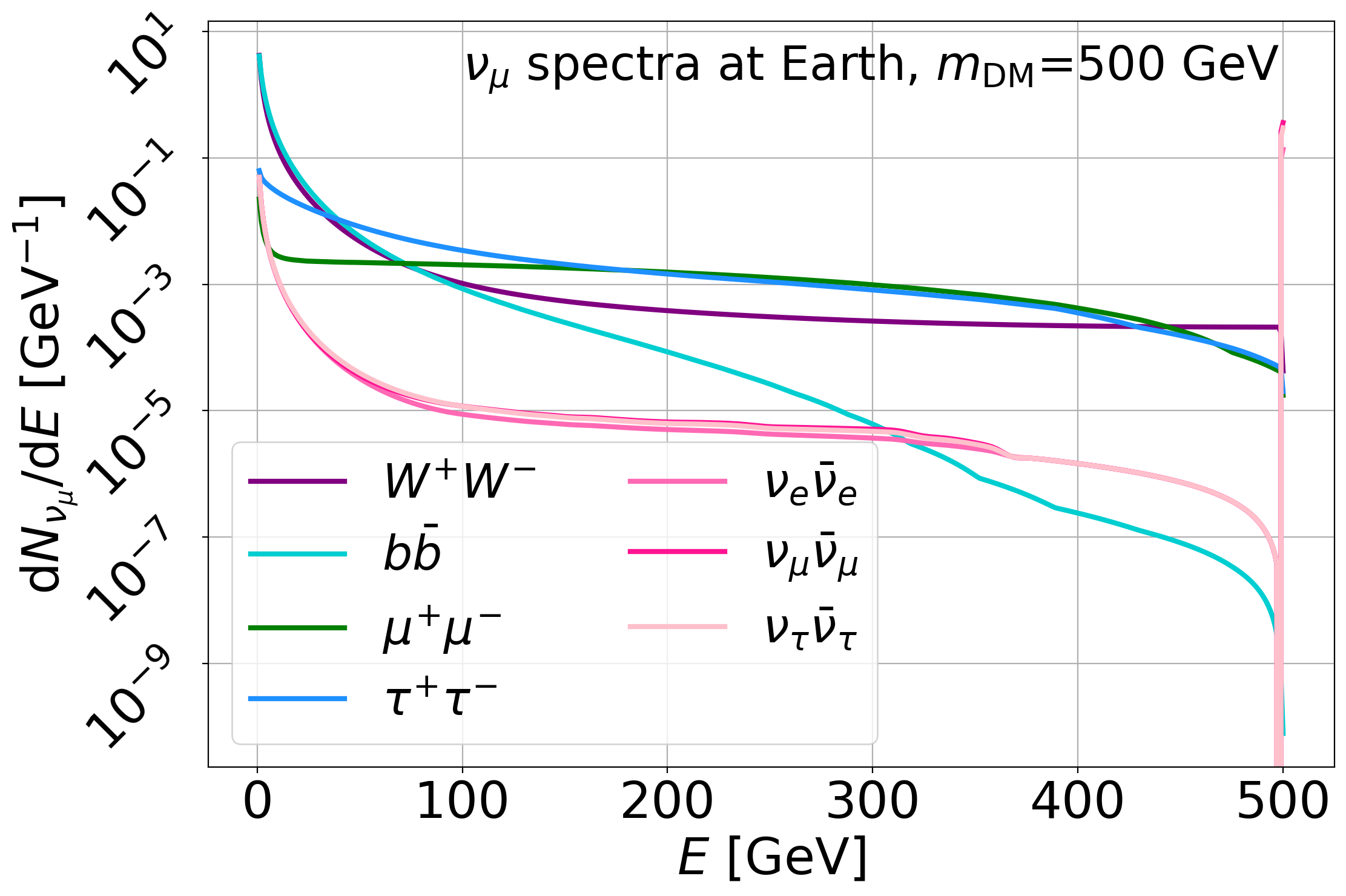}
    \includegraphics[width=0.48\textwidth]{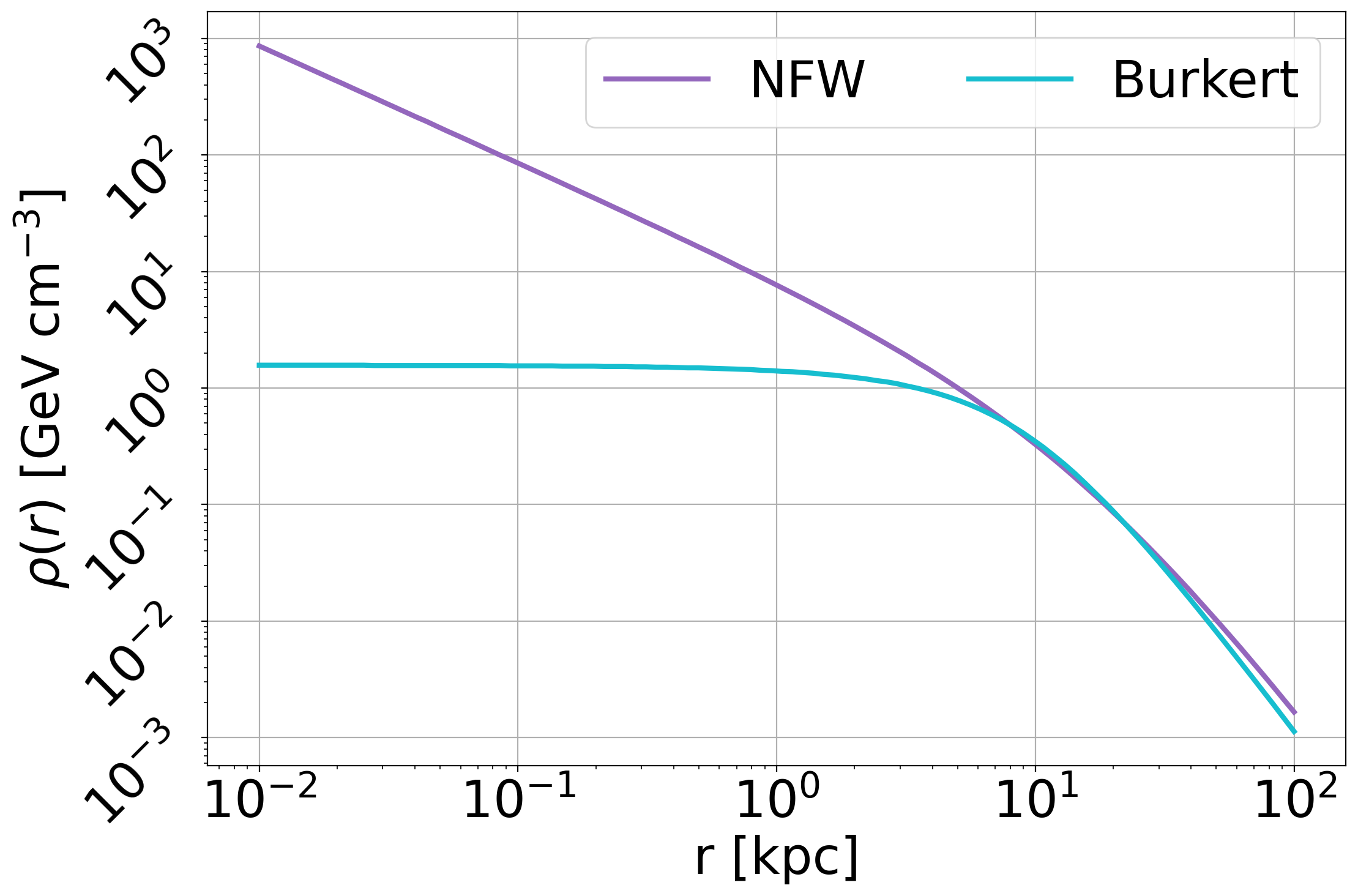}
    \caption{\textbf{Left:} Differential number of muon neutrinos per annihilation expected at Earth for a dark matter mass of 500 GeV. In the analysis, all neutrino flavors are used. \textbf{Right:} Dark matter mass density as a function of the distance to the Galactic Centre for the NFW and Burkert profiles.}
    \label{fig:spectra_profile}
\end{figure}

In this analysis, the neutrino spectra, $\mathrm{d}N_{\mathrm{\nu}}/\mathrm{d}E_{\mathrm{\nu}}$, from DM annihilation/decay, are computed with $\chi a r o \nu $ \cite{Liu_2020}, a framework that couples the PYTHIA simulation with the most up-to-date electroweak correction \cite{Bauer:2020jay} to obtain the neutrino fluxes from dark matter decay and annihilation. These spectra are then propagated to Earth, assuming average oscillation through a very long distance with the oscillation parameters taken from \cite{Esteban:2020cvm}. Figure \ref{fig:spectra_profile} depicts an example of the muon neutrino spectra with a DM mass of 500 GeV for DM annihilation into all of the considered primary channels. The case of DM annihilates/decay directly into neutrinos ($\nu_e \overline{\nu}_e, \nu_\mu \overline{\nu}_\mu, \nu_\tau \overline{\nu}_\tau$) is usually called the \textit{neutrino line} as the spectra contain a sharp peak at $E_{\nu} = m_{DM}$ for annihilation and $E_{\nu} = m_{DM}/2$ for decay as can be seen in the figure. This monochromatic peak is expected from the kinematic of the annihilation or decay while the low-energy tail of the spectra is due to electroweak corrections. Since the sharp peak feature is distinctive from astrophysical backgrounds, the neutrino line channels are expected to yield the best sensitivity for the indirect DM search with neutrino telescopes.

The J-factor is defined as the integration of DM mass density along the line-of-sight and over the field of view represented as a solid angle $\Delta \Omega$:
\begin{equation}
    J(\Psi) = \int_{\Delta \Omega} \int_{0}^{l_{max}}  \rho_{\mathrm{DM}}^{\alpha}(r(l,\Psi)) \, \mathrm{d}l \mathrm{d} \Omega \, ,
\end{equation}
where the dark matter mass density $ \rho_{\mathrm{DM}}$ is assumed to be spherical symmetric such that the value only depends on the distance $r$ from the Galactic Centre, which then can be computed from $\Psi$: the opening angle to the Galactic Center and $l$: the distance along the line-of-sight. The upper integration limit $l_{max}$ is determined based on the radius of the Galactic Halo. For the case of DM annihilation $\alpha = 2$, while for DM decay $\alpha = 1$, where the J-factor is now also referred to as the D-factor.

The actual shape of the DM density profile in the halos of galaxies $\rho_{\mathrm{DM}}$ is still subject to controversy and significant uncertainties \cite{Benito_2019}. There are two main classes of halo shapes regarding whether it exhibits a flat distribution near the core or an exponentially increasing concentration towards the center. The subject is often referred to as the cores-cusp problem \cite{deBlok:2009sp}. For this analysis, we consider two corresponding parametric profiles: Burkert \cite{Burkert_1995} and Navarro-Frenk-White (NFW) \cite{NFW_1996}. The parameter values for these profiles in the case of the Milky Way are adopted from \cite{Nesti:2013uwa}, where they are derived based on motion data within the Galactic region. The mass density of the two profiles is illustrated in Figure \ref{fig:spectra_profile}. The NFW profile represents the `cuspy' profile with the increasing density toward the Galactic Centre and thus is expected to give more optimistic sensitivities compared to the Burkert profile with the flat `core' feature. The Clumpy package \cite{Hutten:2018aix} is used for computing the J-factor as a function of the open angle to the Galactic Centre.

\section{Analysis method}
In this analysis, we use a binned likelihood method on two observables: the reconstructed energy ($E_{reco}$) and the reconstructed open angle of the events ($\psi_{reco}$) to the Galactic Centre. The standard Poisson likelihood function is defined as a product of Poisson probabilities:
\begin{equation}
     \mathcal{L_{\mathrm{Poisson}}}(\xi) = \prod\limits_{i=min}^{max} \text{Poisson} (n_{obs}^{i}, n_{obs}^{tot} f(i ; \xi)) \, ,
\end{equation}
where $n_{obs}^{i}$ is the observed number of events in a given bin $i$. $n_{obs}^{tot}$ is the total number of observed events. $f(i; \xi)$ is the fraction of events in bin $i$ with an assumed hypothesis so that $n_{obs}^{tot} f(i; \xi)$ represents the predicted observation under a hypothesis, but normalized to the total number of observed data. This fraction $f(i; \xi)$ is assumed as followed:
\begin{equation}
     f(i \, ;  \, \xi) = \xi \, \mathcal{S}_i \, + \, (1 - \xi) \mathcal{B}_i \, ,
\end{equation}
with $\mathcal{S}$ and $\mathcal{B}$ being, respectively, the signal and the background probability density functions (PDFs). The only parameter of maximization is the signal fraction $\xi$.

To compute the signal PDF, we use the oscNext Monte Carlo (MC) sample to build a response matrix that encodes how an event with a given true information (flavor, polarization, energy, and open-angle) can interact, being triggered, pass the event selection, and being reconstructed. A kernel density estimation (KDE) is also applied to construct a smooth response matrix. This matrix is then used to convolute the true flux in Equation (\ref{eq:trueflux}) into the expected signal event distribution. For each combination of channels, density profile, and annihilation/decay process, we only consider the dark matter mass range such that the median of the signal PDF is within  95\% upper and lower bound of the reconstruction energy of the MC simulation. The reason is that signal PDFs that do not fulfill these criteria will mostly use MC regions with a lack of statistics, which leads to unreliable PDF shapes. Additionally, these signal PDFs peak at the energy region of which most of the events will be filtered out by the selection.

The background PDF is estimated from right-ascension scrambled data which is a widely used method for background estimation in the neutrino telescope. In this approach, each data event is assigned with a random right-ascension coordinate to make a pseudo-sample following the background hypothesis. The background PDF is then estimated as an average of 100 right-ascension scrambled pseudo-samples with the same KDE technique applied as for the case of signal PDF. The scrambling method is valid due to a detector up-time of $>98\%$ for the telescope and the rotation of the Earth, which makes the background of atmospheric neutrinos and muons to be uniform in right-ascension.

The scrambling technique is powerful such that it can account for any systematic uncertainties that could affect the background model. Nevertheless, if there is a possible signal in the data, it could potentially contaminate the data-scrambling estimation of the background. To correct for that, the signal subtraction likelihood \cite{IceCube:2016oqp} is used such that the estimation of background-only PDF reads:
\begin{equation}
    \mathcal{B}_i = \frac{1}{1-\xi} (\mathcal{B}_i^{\mathrm{scrambled}} - \xi \mathcal{S}^{\mathrm{scrambled}}_i) \, ,
\end{equation}
with $\mathcal{S}^{\mathrm{scrambled}}_i$ being the right ascension scrambled signal PDF. The final form of the hypothesis PDF now yields:

\begin{equation}
     f(i;\xi) = \xi \mathcal{S}_i  + \mathcal{B}^{\mathrm{scrambled}}_i - \xi \mathcal{S}^{\mathrm{scrambled}}_i  \, .
\end{equation}

\begin{figure}
    \centering
    \includegraphics[width=0.3\textwidth]{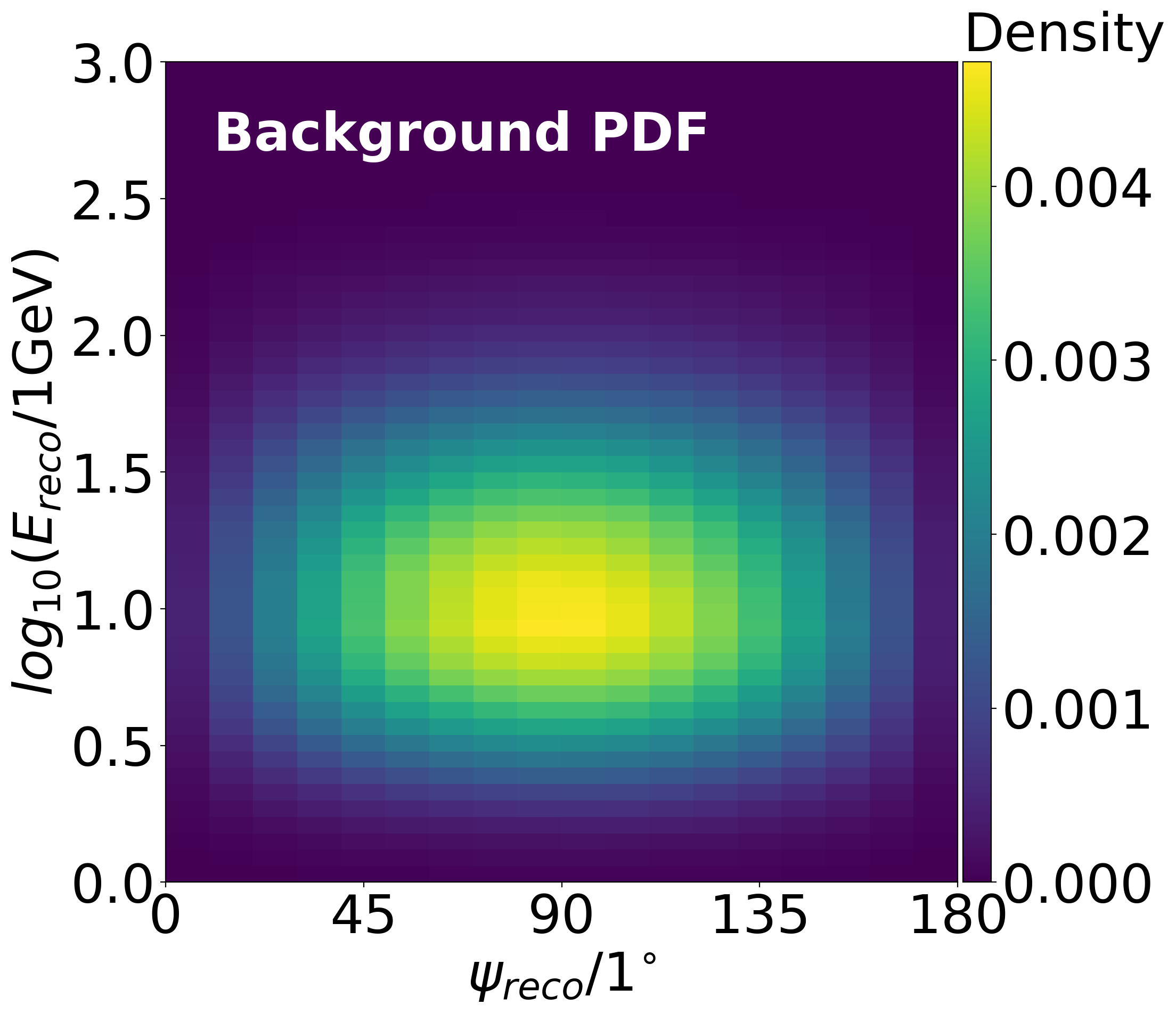}    \includegraphics[width=0.3\textwidth]{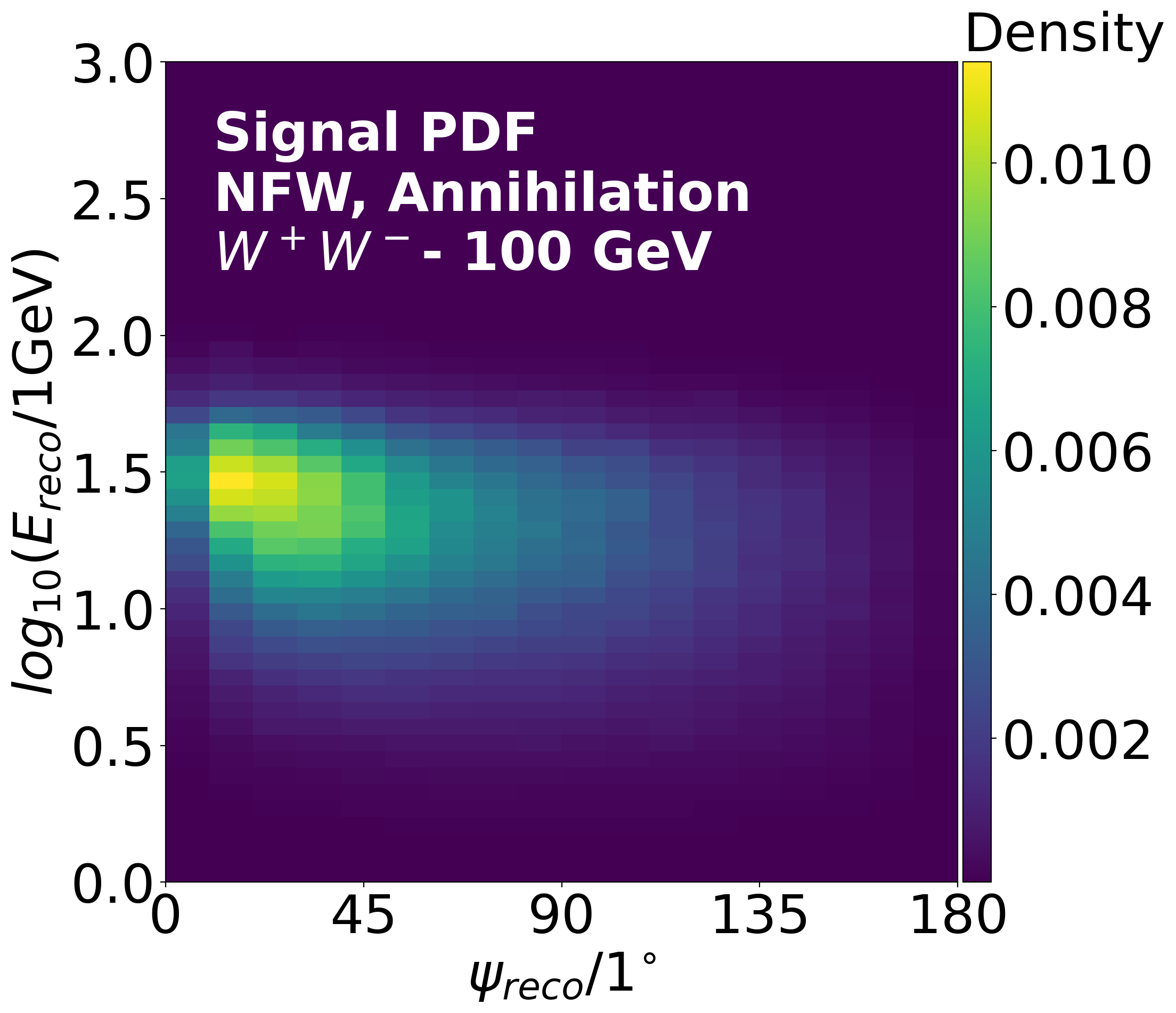}
    \includegraphics[width=0.3\textwidth]{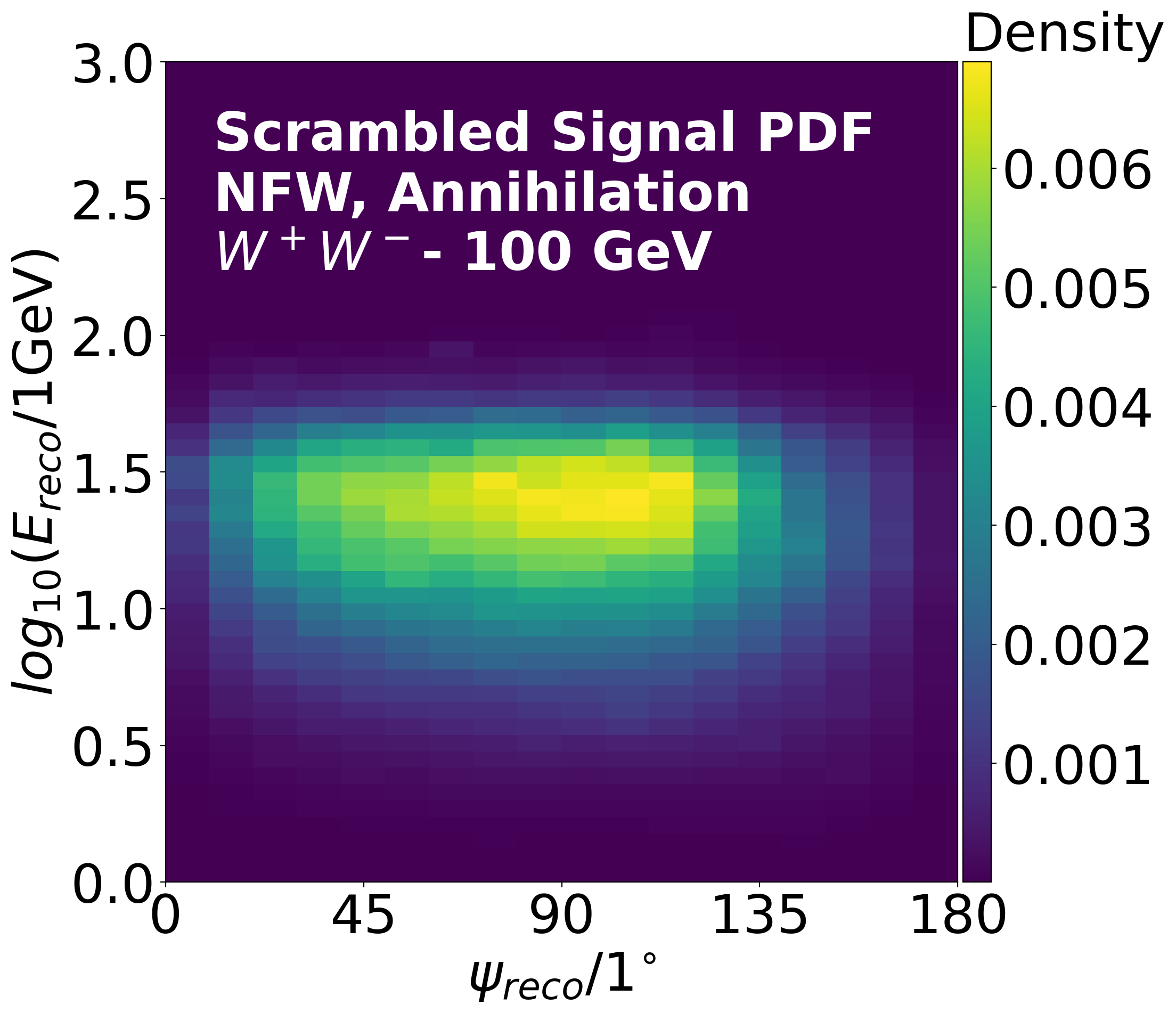}
    \caption{Two-dimensional PDFs on reconstruction energy ($E_{reco}$) and reconstruction open angle to the GC ($\psi_{reco}$) for: background from right-ascension scrambled data \textit{(left)}, signal of DM with mass 100 GeV annihilating into $W^+ W^-$ assuming NFW profile \textit{(middle)}, and the same signal but scrambled in right ascension \textit{(right)}.}
    \label{fig:PDFs}
\end{figure}

Figure \ref{fig:PDFs} illustrates these PDFs with an example of signal in the case of DM with mass 100 GeV annihilating into $W^+W^-$, and the DM profile is assumed to be NFW. One can see that the signal can be distinguished from the background such that it is more pronounced in the Galactic Centre region.

\section{Sensitivity and Conclusions}
If the best-fit signal fraction $\xi$ is consistent with the background hypothesis, one can then derive an upper limit on the signal fraction at $90\%$ confidence level (CL): $\xi_{90}$. In this work, we make use of the log-likelihood interval method \cite{Cowan:2010js} for the computation of $\xi_{90}$. For each signal combination of mass, profile (NFW/Burkert) and process (annihilation/decay), 1000 pseudo-samples are generated as the Poissonian variation of the background-only hypothesis. The median value of $\xi_{90}$ derived from each of these samples is quoted as the 90\% CL median sensitivity. As indicated in Equation (\ref{eq:trueflux}), the total number of signal events is proportional to the decay lifetime $\tau$ or the velocity-averaged cross-section $\langle \sigma \upsilon \rangle$. Thus, the limits on signal fraction $\xi_{90}$ can be converted into the limit on these physics parameters.

\begin{figure}
    \centering
    \includegraphics[width=0.4 \textwidth]{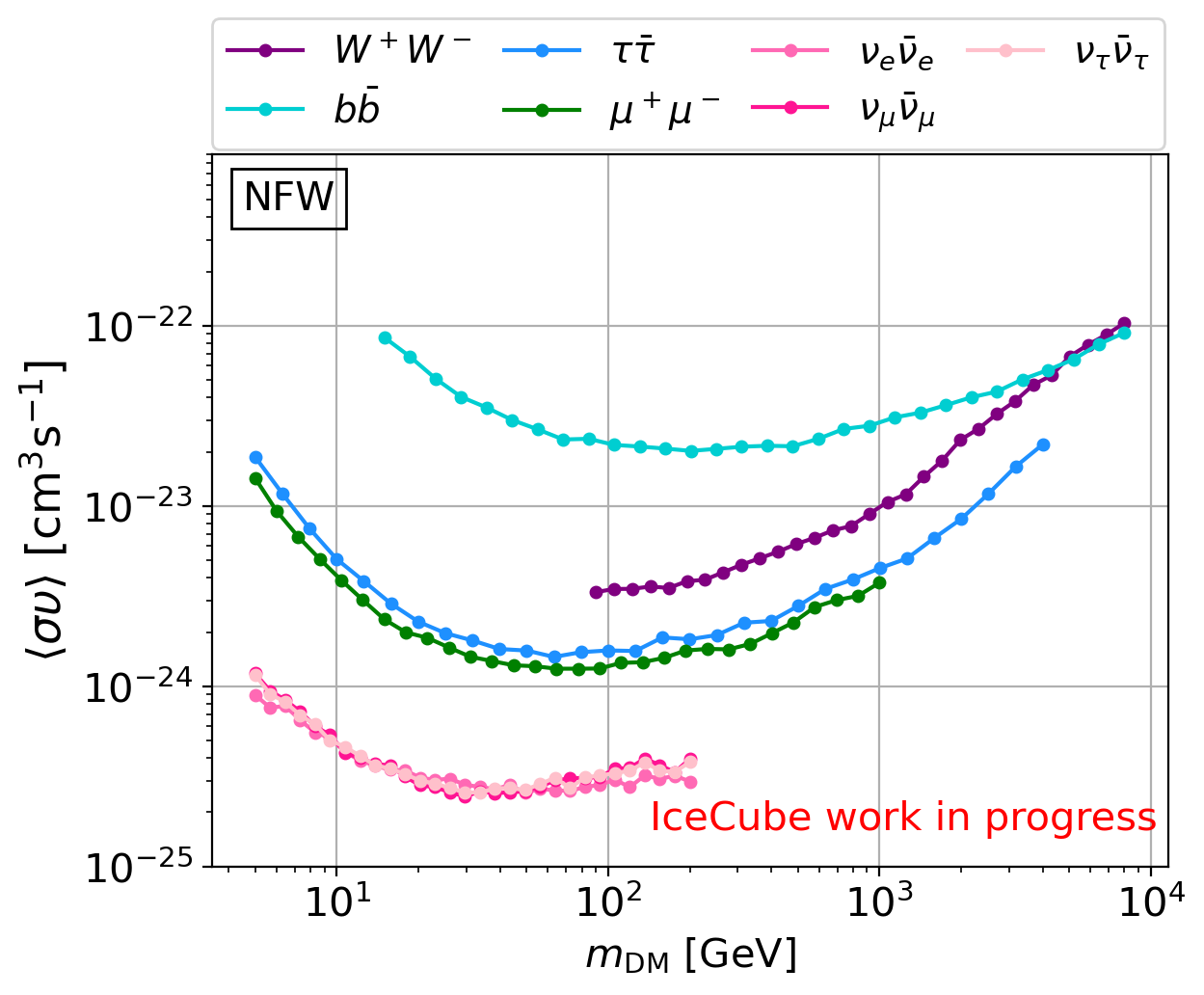}
    \includegraphics[width=0.4 \textwidth]{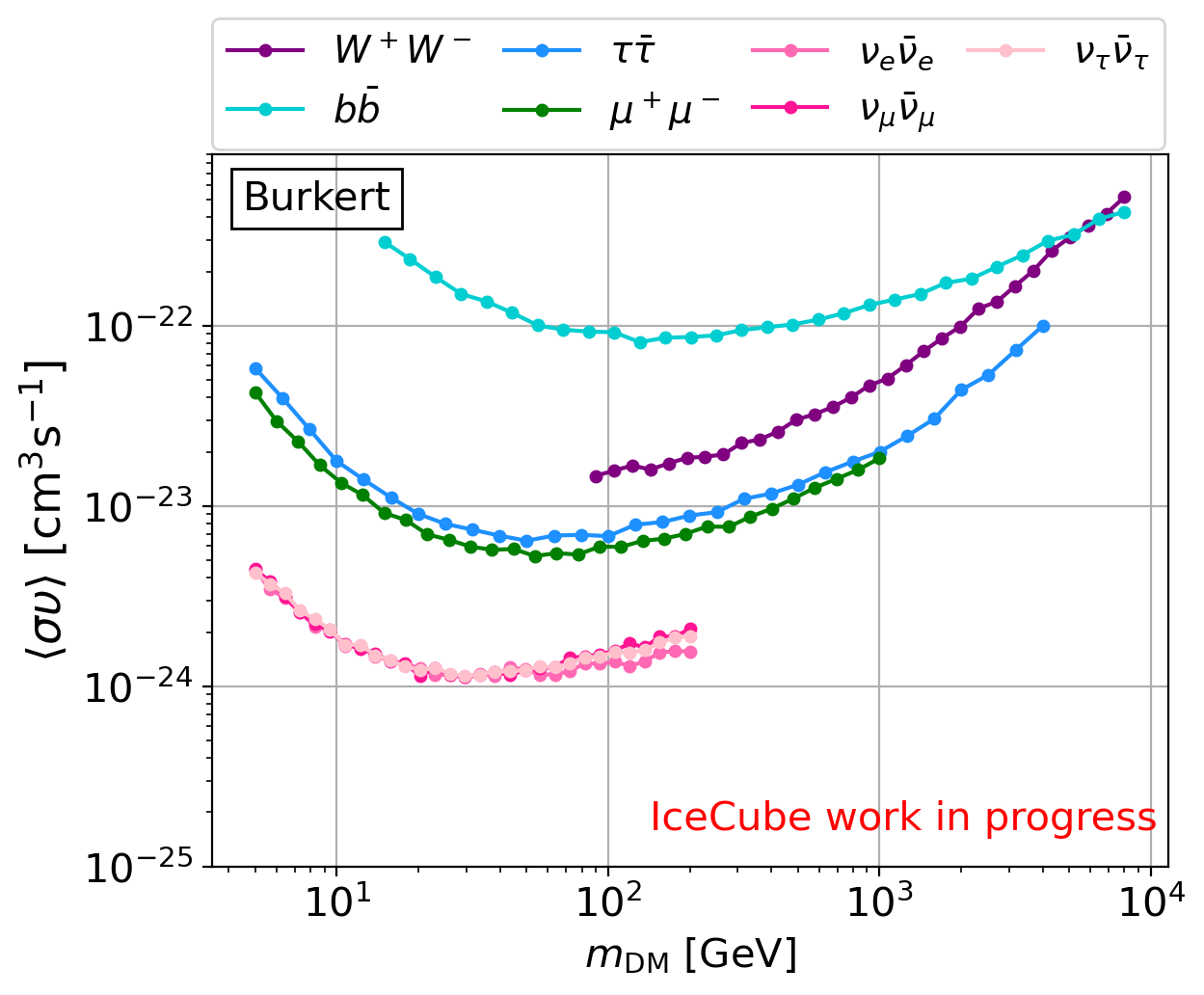}
    \includegraphics[width=0.4 \textwidth]{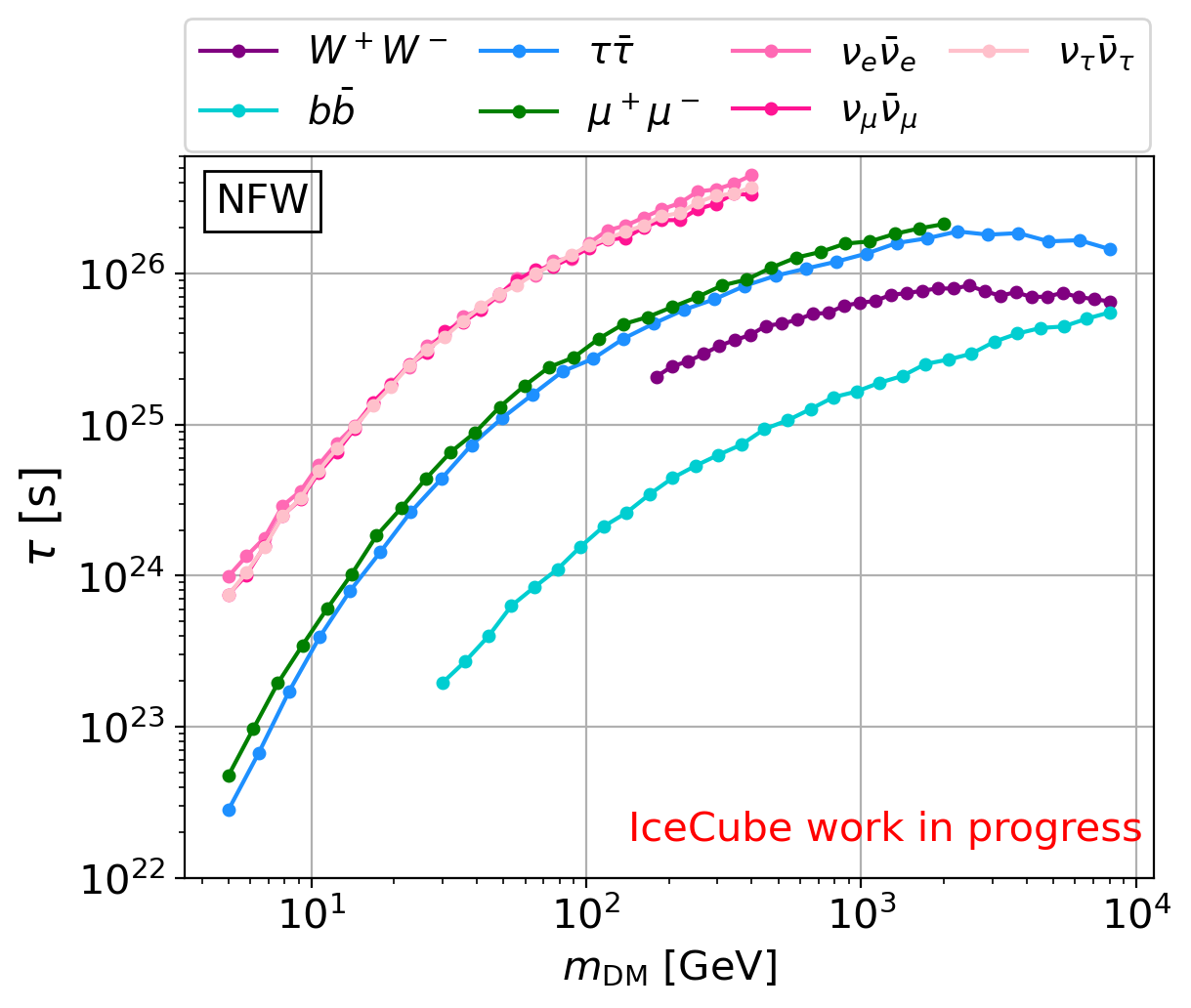}
    \includegraphics[width=0.4 \textwidth]{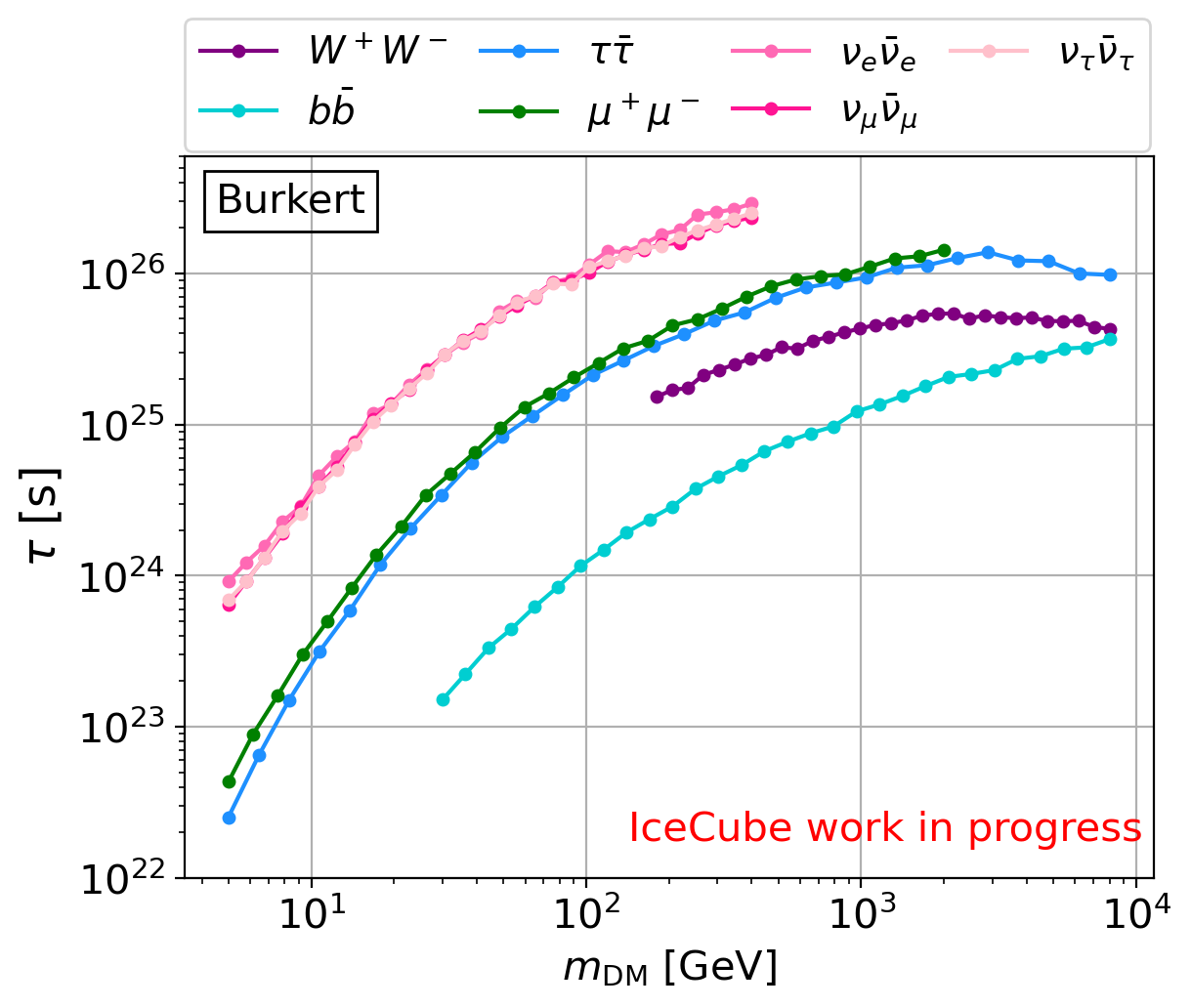}
    \caption{Sensitivities on the thermally-averaged dark matter self-annihilation cross-section $\langle \sigma \upsilon \rangle$ \textit{(upper)} and dark matter lifetime $\tau$ \textit{(lower)} as a function of the dark matter mass for both the NFW \textit{(left)} and Burkert \textit{(right)} halo profiles.}
    \label{fig:Limits}
\end{figure}

\begin{figure}
    \centering
    \includegraphics[width=0.5 \textwidth]{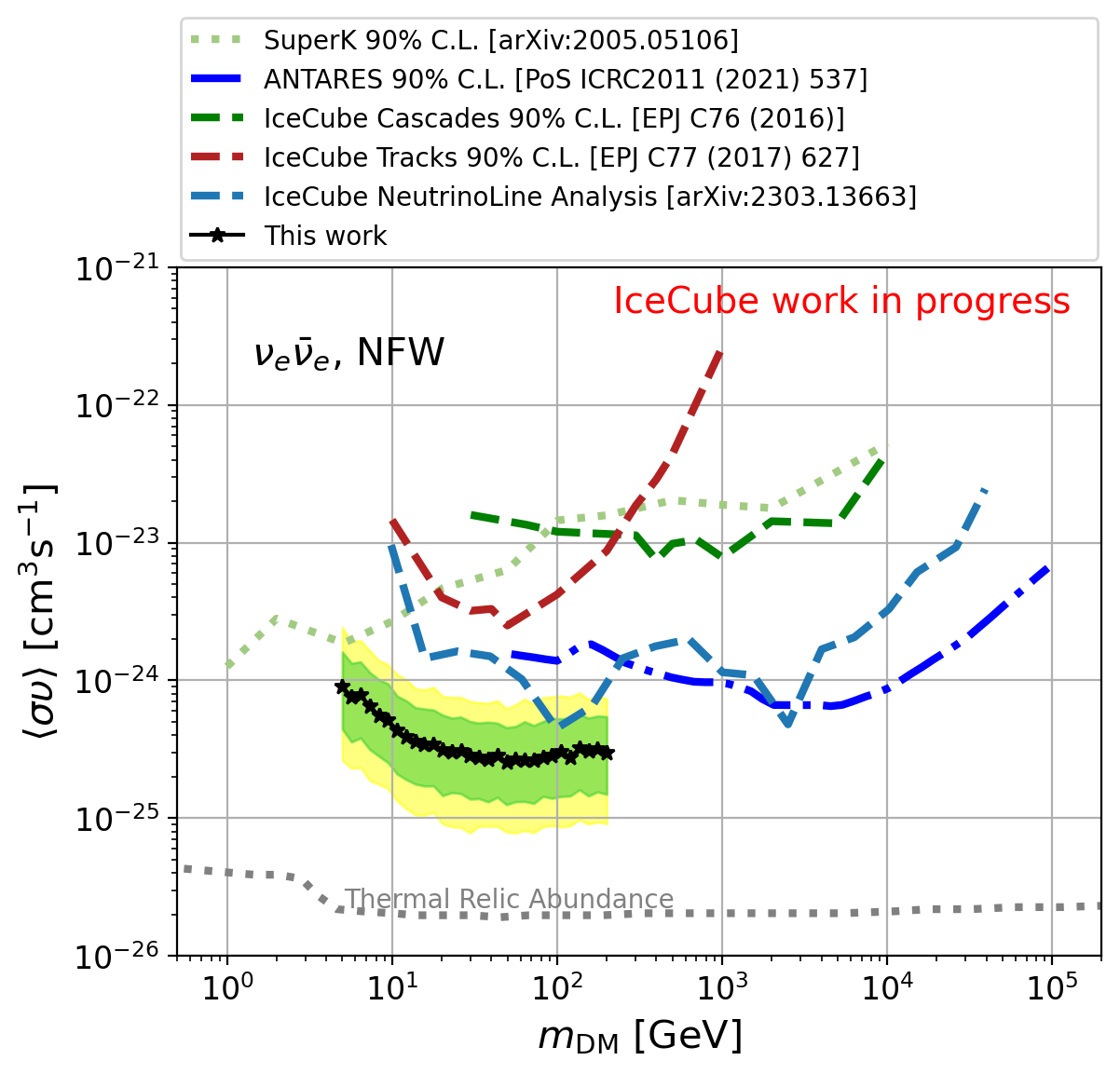}
    \caption{Sensitivity on the thermally averaged cross-section for the $\nu_e \overline{\nu}_e$ channels (green and yellow represent 1 and 2$\sigma$ bands) compared to previous IceCube results \cite{IceCube:2017rdn, IceCube:2023ies, IceCube:2016oqp}, as well as Super-Kamiokande \cite{Super-Kamiokande:2020sgt} and ANTARES \cite{KM3NeT:2021kbg}.The dotted grey line is the cross-section required to produce the observed relic abundance from thermal freeze-out computed in \cite{Steigman:2012nb}.}
    \label{fig:Limits_Comparison}
\end{figure}

In Figure \ref{fig:Limits}, we present the sensitivity as the median upper/lower limit at the 90\% CL on velocity-averaged annihilation cross-section $\langle \sigma \upsilon \rangle$ and DM lifetime $\tau$ respectively. A comparison with other experiments and previous IceCube analyses in terms of the $\nu_e\overline{\nu}_e$ neutrino line channel for annihilation and NFW profile is presented in Figure \ref{fig:Limits_Comparison}. One can expect an improvement compared to the latest result on the neutrino line search with IceCube \cite{IceCube:2023ies}. This enhancement comes from more years of data included as well as substantial development of the data selection, which specifically targets the low-energy region. The final official results of this work will be available soon.

\clearpage
\bibliographystyle{ICRC}
\bibliography{ICRC2023_template_IceCube}

%

\clearpage

\section*{Full Author List: IceCube Collaboration}

\scriptsize
\noindent
R. Abbasi$^{17}$,
M. Ackermann$^{63}$,
J. Adams$^{18}$,
S. K. Agarwalla$^{40,\: 64}$,
J. A. Aguilar$^{12}$,
M. Ahlers$^{22}$,
J.M. Alameddine$^{23}$,
N. M. Amin$^{44}$,
K. Andeen$^{42}$,
G. Anton$^{26}$
% ...
\\
\\
% ...
$^{12}$ Universit{\'e} Libre de Bruxelles, Science Faculty CP230, B-1050 Brussels, Belgium \\
% ...
$^{17}$ Department of Physics, Loyola University Chicago, Chicago, IL 60660, USA \\
$^{18}$ Dept. of Physics and Astronomy, University of Canterbury, Private Bag 4800, Christchurch, New Zealand \\
% ...
$^{22}$ Niels Bohr Institute, University of Copenhagen, DK-2100 Copenhagen, Denmark \\
$^{23}$ Dept. of Physics, TU Dortmund University, D-44221 Dortmund, Germany \\
% ...
$^{26}$ Erlangen Centre for Astroparticle Physics, Friedrich-Alexander-Universit{\"a}t Erlangen-N{\"u}rnberg, D-91058 Erlangen, Germany \\
% ...
$^{40}$ Dept. of Physics and Wisconsin IceCube Particle Astrophysics Center, University of Wisconsin{\textendash}Madison, Madison, WI 53706, USA \\
% ...
$^{42}$ Department of Physics, Marquette University, Milwaukee, WI, 53201, USA \\
% ...
$^{44}$ Bartol Research Institute and Dept. of Physics and Astronomy, University of Delaware, Newark, DE 19716, USA \\
% ...
$^{63}$ Deutsches Elektronen-Synchrotron DESY, Platanenallee 6, 15738 Zeuthen, Germany  \\
$^{64}$ Institute of Physics, Sachivalaya Marg, Sainik School Post, Bhubaneswar 751005, India \\
% ...

\subsection*{Acknowledgements}

\noindent
The authors gratefully acknowledge the support from the following agencies and institutions:
USA {\textendash} U.S. National Science Foundation-Office of Polar Programs,
U.S. National Science Foundation-Physics Division,
U.S. National Science Foundation-EPSCoR,
Wisconsin Alumni Research Foundation,
Center for High Throughput Computing (CHTC) at the University of Wisconsin{\textendash}Madison,
Open Science Grid (OSG),
Advanced Cyberinfrastructure Coordination Ecosystem: Services {\&} Support (ACCESS),
Frontera computing project at the Texas Advanced Computing Center,
U.S. Department of Energy-National Energy Research Scientific Computing Center,
Particle astrophysics research computing center at the University of Maryland,
Institute for Cyber-Enabled Research at Michigan State University,
and Astroparticle physics computational facility at Marquette University;
Belgium {\textendash} Funds for Scientific Research (FRS-FNRS and FWO),
FWO Odysseus and Big Science programmes,
and Belgian Federal Science Policy Office (Belspo);
Germany {\textendash} Bundesministerium f{\"u}r Bildung und Forschung (BMBF),
Deutsche Forschungsgemeinschaft (DFG),
Helmholtz Alliance for Astroparticle Physics (HAP),
Initiative and Networking Fund of the Helmholtz Association,
Deutsches Elektronen Synchrotron (DESY),
and High Performance Computing cluster of the RWTH Aachen;
Sweden {\textendash} Swedish Research Council,
Swedish Polar Research Secretariat,
Swedish National Infrastructure for Computing (SNIC),
and Knut and Alice Wallenberg Foundation;
European Union {\textendash} EGI Advanced Computing for research;
Australia {\textendash} Australian Research Council;
Canada {\textendash} Natural Sciences and Engineering Research Council of Canada,
Calcul Qu{\'e}bec, Compute Ontario, Canada Foundation for Innovation, WestGrid, and Compute Canada;
Denmark {\textendash} Villum Fonden, Carlsberg Foundation, and European Commission;
New Zealand {\textendash} Marsden Fund;
Japan {\textendash} Japan Society for Promotion of Science (JSPS)
and Institute for Global Prominent Research (IGPR) of Chiba University;
Korea {\textendash} National Research Foundation of Korea (NRF);
Switzerland {\textendash} Swiss National Science Foundation (SNSF);
United Kingdom {\textendash} Department of Physics, University of Oxford.

\end{document}